# Novel X band Compact Waveguide Dual Circular Polarizer

Chen Xu, Sami Tantawi, Juwen Wang

SLAC National Accelerator Laboratory, Stanford University, Stanford, CA 94309, USA
Electronic address: chenxu@slac.stanford.edu

Abstract

A novel type of dual circular polarizer is developed to convert the TE10 mode into two different polarization TE11 modes in a circular waveguide. This design consists two major parts: a TE10 to TE10/TE20 converter and an overmoded TE10/TE20 to circular TE11 modes converter.

## 1. Introduction

Cosmist microwave background (CMB) polarization experiment is required to understand the basics of early universe by using X band to W band microwave. An orthomode transducer (OMT) attached to the horn antenna is used to separate two orthogonally polarized microwave signals into TE10 and TE01 modes into two different rectangular waveguides. To study the amplitudes and phase of these TE10 or TE01 modes, one can learn the RF characteristics of the emitters, such as planets. On the contrary, from the broadcast RF wave point of view, one can combine two TE10 modes in different rectangular waveguide into polarized modes in a circular waveguide. A klystron can deliver RF power in form of TE10 mode in a rectangular mode, it will be convenient for a converter to convert TE10 mode into two TE11 modes with 90 degree phase difference in a circular waveguide. An orthmode transducer with four arm waveguides is recently introduced. This turnstile junction structure has four rectangular ports which support TE10 mode. On the top of this junction, there is a circular waveguide which outputs two circular polarized TE11 modes. We want to design a converter which has two rectangular ports and one circular waveguide, and the input power from first rectangular waveguide will be split into two TE11 polarization modes with 90 degree in a circular waveguide. This combined signal is named right hand circular polarized. While a signal received is right hand circular polarized, it would output solely to the second rectangular waveguide on the other side. The diagram of power flow is illustrated in Fig.1.

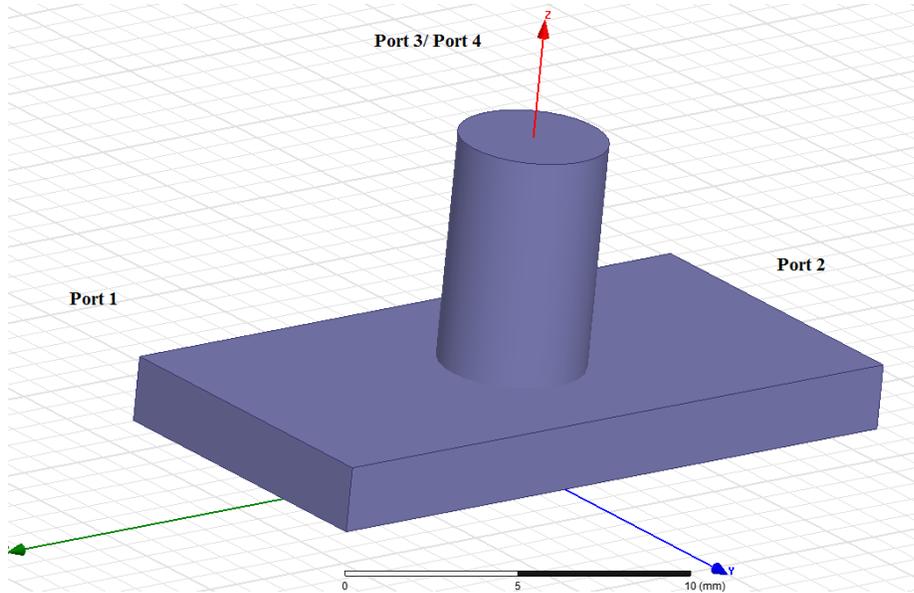

Fig.1 : Demonstration design of circular polarizer and its power flow.

## 2. Design concept

This device has three physical ports but the circular waveport supports two polarizations, thus it is a four ports structure with a 4 by 4 scattering matrix. To achieve the goal that we describe in the Fig.1, the targeted scattering matrix for these four modes should be the following:

$$S = \frac{1}{2} \times \begin{bmatrix} 0 & 0 & \sqrt{2} & -\sqrt{2}i \\ 0 & 0 & -\sqrt{2} & -\sqrt{2}i \\ \sqrt{2} & -\sqrt{2} & 0 & 0 \\ -\sqrt{2}i & -\sqrt{2}i & 0 & 0 \end{bmatrix}$$

When the RF power comes in from the port1 and splits equally into the mode 1 and the mode 2 in the third port, and those two modes have 90 degree phase difference. If the RF power comes from port 2 and it will split equally into those two modes with 270 degree delay. However, when the power received from port 3 with two modes apart 180 degree phase difference will produce distribute the RF power equally on two rectangular ports.

However, the matrix above is oversimplified, one needs to cascade the phase shift matrix with the center scattering matrix and the phase shift matrix is based on the lengths of input and output arm waveguide. Thus, there is more general form of this 4 by 4 matrix, it is shown as following.

$$S = \frac{1}{2} \times \begin{bmatrix} 0 & 0 & \sqrt{2} \times e^{-j\omega \frac{L_1}{\beta_1}} & -\sqrt{2}i \times e^{-j\omega \frac{L_2}{\beta_2}} \\ 0 & 0 & -\sqrt{2} \times e^{-j\omega \frac{L_1}{\beta_1}} & -\sqrt{2}i \times e^{-j\omega \frac{L_2}{\beta_2}} \\ \sqrt{2} \times e^{-j\omega \frac{L_1}{\beta_1}} & -\sqrt{2} \times e^{-j\omega \frac{L_1}{\beta_1}} & 0 \times e^{-j\omega \frac{L_3}{\beta_3}} & 0 \\ -\sqrt{2}i \times e^{-j\omega \frac{L_2}{\beta_2}} & -\sqrt{2}i \times e^{-j\omega \frac{L_2}{\beta_2}} & 0 & 0 \times e^{-j\omega \frac{L_3}{\beta_3}} \end{bmatrix}$$

Where $L_1$, $L_2$ and $L_3$ are the length of three arms and $\beta_1$ $\beta_2$ and $\beta_3$ are the propagation constants in each waveguide. In our case, $\beta_1$ and $\beta_2$ are the same.

How does TE10 mode convert into a circular waveguide? In Fig.2, a rectangular waveguide that supports TE10 mode, the field pattern shown that the H field is looping around in the longitude direction. This H field will be easily coupled to a TE11 mode in circular waveguide. Moreover, the rectangular waveguide also supports TE20 mode and the H field has two loops in the transverse direction. This H field can easily couple to another polarization TE11 mode in the circular waveguide.

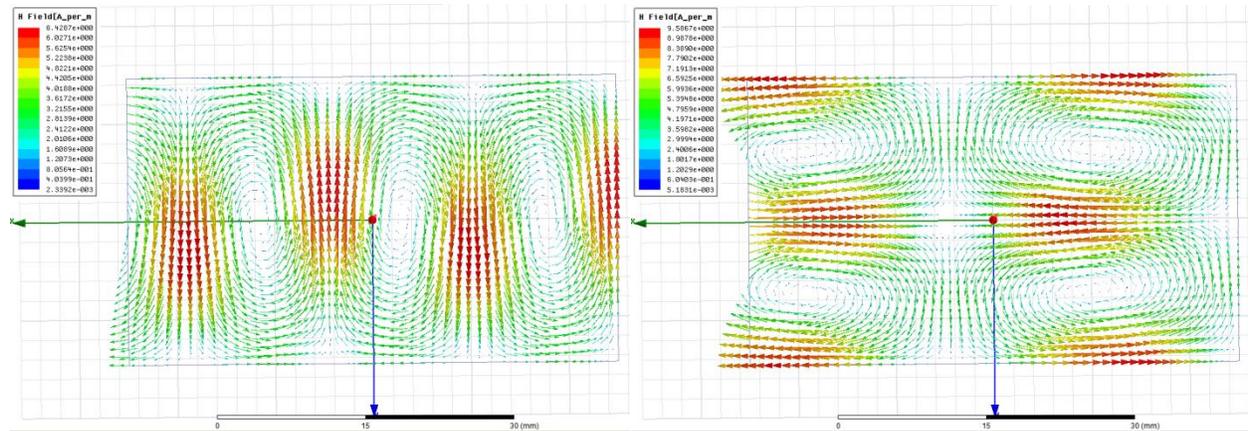

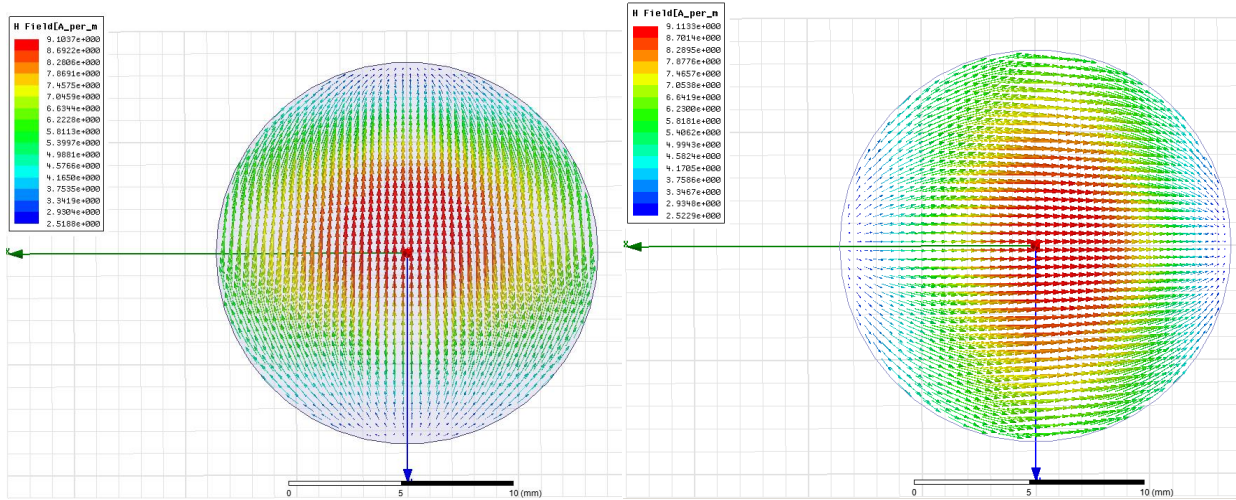

Fig. 2. Demonstration on vector magnetic field patterns of TE10 and TE20 in a rectangular waveguide and TE11 modes in a circular waveguide.

To obtain a combination of these two scenarios simultaneously, an overmoded rectangular waveguide is utilized to flow TE10 and TE20 modes with equal power. If the coupling indexes to two TE11 modes are equal, the two TE11 modes have the same power. The phase delay of these two TE11 modes are determined by the phase delay of two modes in rectangular waveguide. Note, these two modes in rectangular waveguide have different propagation constants, meaning that the guided wavelengths of two rectangular modes are different. With such consideration above, the center unit has three physical ports and each of ports supports two modes. Thus, the scattering matrix is expending into a 6-by-6 matrix and this unitary matrix is shown:

$$S = \begin{bmatrix} \dfrac{-\cos[\theta_{01}] - e^{i\phi_{01}}}{2} & 0 & \dfrac{-\cos[\theta_{01}] + e^{i\phi_{01}}}{2} & 0 & \dfrac{\sin[\theta_{01}]}{\sqrt{2}} & 0 \\ 0 & \dfrac{-\cos[\theta_{02}] - e^{i\phi_{02}}}{2} & 0 & \dfrac{-\cos[\theta_{02}] + e^{i\phi_{02}}}{2} & 0 & \dfrac{\sin[\theta_{02}]}{\sqrt{2}} \\ \dfrac{-\cos[\theta_{01}] + e^{i\phi_{01}}}{2} & 0 & \dfrac{-\cos[\theta_{01}] - e^{i\phi_{01}}}{2} & 0 & \dfrac{\sin[\theta_{01}]}{\sqrt{2}} & 0 \\ 0 & \dfrac{-\cos[\theta_{02}] + e^{i\phi_{02}}}{2} & 0 & \dfrac{-\cos[\theta_{02}] - e^{i\phi_{02}}}{2} & 0 & \dfrac{\sin[\theta_{02}]}{\sqrt{2}} \\ \dfrac{\sin[\theta_{01}]}{\sqrt{2}} & 0 & \dfrac{\sin[\theta_{01}]}{\sqrt{2}} & 0 & \cos[\theta_{01}] & 0 \\ 0 & \dfrac{\sin[\theta_{02}]}{\sqrt{2}} & 0 & \dfrac{\sin[\theta_{02}]}{\sqrt{2}} & 0 & \cos[\theta_{02}] \end{bmatrix}$$

This matrix is unitary and $\theta_{01}$ $\theta_{02}$ $\phi_{01}$ and $\phi_{02}$ are optimization parameters. One can simplify this matrix by equaling $\theta_{01}$ and $\theta_{02}$ and equaling $\phi_{01}$ and $\phi_{02}$.

The matrix is regressive to

$$S = \begin{bmatrix} \dfrac{-\cos[\theta] - e^{i\phi_{01}}}{2} & 0 & \dfrac{-\cos[\theta] + e^{i\phi}}{2} & 0 & \dfrac{\sin[\theta]}{\sqrt{2}} & 0 \\ 0 & \dfrac{-\cos[\theta] - e^{i\phi}}{2} & 0 & \dfrac{-\cos[\theta] + e^{i\phi}}{2} & 0 & \dfrac{\sin[\theta]}{\sqrt{2}} \\ \dfrac{-\cos[\theta] + e^{i\phi}}{2} & 0 & \dfrac{-\cos[\theta] - e^{i\phi}}{2} & 0 & \dfrac{\sin[\theta]}{\sqrt{2}} & 0 \\ 0 & \dfrac{-\cos[\theta] + e^{i\phi}}{2} & 0 & \dfrac{-\cos[\theta] - e^{i\phi}}{2} & 0 & \dfrac{\sin[\theta]}{\sqrt{2}} \\ \dfrac{\sin[\theta]}{\sqrt{2}} & 0 & \dfrac{\sin[\theta]}{\sqrt{2}} & 0 & \cos[\theta] & 0 \\ 0 & \dfrac{\sin[\theta]}{\sqrt{2}} & 0 & \dfrac{\sin[\theta]}{\sqrt{2}} & 0 & \cos[\theta] \end{bmatrix}$$

Because the modes in the port 1 and port 2 are isolated, in a steady state, there are two partial standing waves inside this center unit and the peak H field is at the center location where the circular waveguide are. For this reason, the first four number in the input and output vector of this matrix should have the same amplitude but 180 degree phase difference. To satisfy this condition, the analytical solution is from Mathematica and shown in following:

$$S = \begin{bmatrix} \frac{1}{2} \times (\frac{1}{3} - e^{i\phi}) & 0 & \frac{1}{2} \times (\frac{1}{3} + e^{i\phi}) & 0 & \frac{2}{3} & 0 \\ 0 & \frac{1}{2} \times (\frac{1}{3} - e^{i\phi}) & 0 & \frac{1}{2} \times (\frac{1}{3} + e^{i\phi}) & 0 & \frac{2i}{3} \\ \frac{1}{2} \times (\frac{1}{3} + e^{i\phi}) & 0 & \frac{1}{2} \times (\frac{1}{3} - e^{i\phi}) & 0 & \frac{2}{3} & 0 \\ 0 & \frac{1}{2} \times (\frac{1}{3} + e^{i\phi}) & 0 & \frac{1}{2} \times (\frac{1}{3} - e^{i\phi}) & 0 & \frac{2i}{3} \\ \frac{2}{3} & 0 & \frac{2}{3} & 0 & -\frac{1}{3} & 0 \\ 0 & \frac{2i}{3} & 0 & \frac{2i}{3} & 0 & \frac{1}{3} \end{bmatrix}$$

This matrix has one degree of freedom $\phi$. If $\phi$ is chosen as 0 degree, the matrix is further simplified:

$$S = \begin{bmatrix} -\frac{1}{3} & 0 & \frac{2}{3} & 0 & \frac{2}{3} & 0 \\ 0 & -\frac{1}{3} & 0 & \frac{2}{3} & 0 & \frac{2i}{3} \\ \frac{2}{3} & 0 & -\frac{1}{3} & 0 & \frac{2}{3} & 0 \\ 0 & \frac{2}{3} & 0 & -\frac{1}{3} & 0 & \frac{2i}{3} \\ \frac{2}{3} & 0 & \frac{2}{3} & 0 & -\frac{1}{3} & 0 \\ 0 & \frac{2i}{3} & 0 & \frac{2i}{3} & 0 & \frac{1}{3} \end{bmatrix}$$

To achieve this goal matrix, we implemented HFSS simulation to obtain the numerical solution. While a circular waveguide is required to only allow TE11 modes propagate at 11.424 Ghz, A wide rectangular waveguide is required to accommodate the TE10 and TE20 modes. TE10 and TE20 modes have different propagate constant in this waveguide. Simply, TE20 has two times propagation constant than that of TE10, and it means that in each TE10 wave package, there are two TE20 wave package if TE10 and TE20 are in phase. This design and its E field pattern on the surface are illustrated in the Fig. 3. The scattering matrix is shown in table 1.

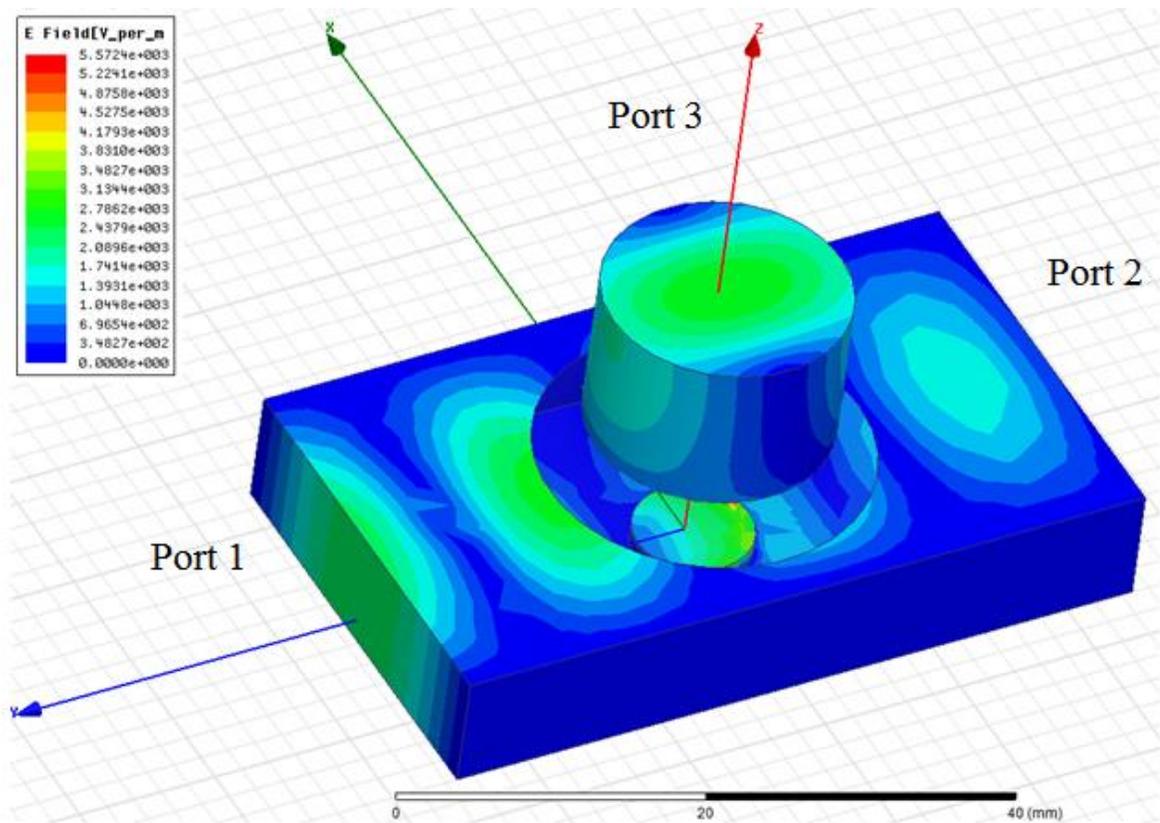

Fig.3: This is the E field pattern on the mode converter. The input power on port1 is 1 W for TE10/TE20 modes.

Port 1 and port 2 are TE10 and TE20 overmoded rectangular waveguide, and port3 are the circular waveguide. On the rectangular waveguide surface, E field pattern are combination of TE10 and TE20, and they are converted to two TE11 modes with different polarization with 90 degree phase difference at the bottom of the circular waveguide. Since the two TE11 have same field amplitude and 90 degree out of phase, the outcome of the port 3 reveals one TE11 rotating around the center. The max E field occurs at the bottom of the circular waveguide.

Table.1: The scattering matrix of device in fig.3. Phase unit is degree.

| Freq | S:1:1 | S:1:2 | S:2:1 | S:2:2 | S:3:1 | S:3:2 |
|---|---|---|---|---|---|---|
| 11.424 (GHz) 1:1 | ( 0.33575, 104) | ( 0.00072212, 149) | ( 0.66333, -78.3) | ( 0.00023088, 147) | ( 0.66878, 175) | ( 0.00056209, 11.3) |
| 1:2 | ( 0.00071615, 149) | ( 0.43163, -131) | ( 0.0003214, -26.7) | ( 0.59732, 21.6) | ( 0.00032535, 61.9) | ( 0.67595, -108) |
| 2:1 | ( 0.66333, -78.3) | ( 0.00031875, -26.3) | ( 0.33782, 104) | ( 0.00082951, -36.4) | ( 0.66773, 175) | ( 0.00063952, -163) |
| 2:2 | ( 0.00022563, 148) | ( 0.59733, 21.6) | ( 0.00083073, -36.4) | ( 0.43169, -131) | ( 0.00023547, -85.3) | ( 0.67591, -108) |
| 3:1 | ( 0.66878, 175) | ( 0.00033048, 61.5) | ( 0.66773, 175) | ( 0.00023332, -84.4) | ( 0.32691, -110) | ( 9.9917e-005, -63.4) |
| 3:2 | ( 0.00056913, 11.4) | ( 0.67595, -108) | ( 0.00062978, -163) | ( 0.6759, -108) | ( 0.00010825, -61.9) | ( 0.29368, -15) |

It is necessary to convert TE10 into TE10 and TE20 overmoded with same and amplitude and arbitrary phase difference. The general scattering matrix is shown and the simplified is following.

$$S = \begin{bmatrix} \frac{\cos[\theta]}{2} & \frac{\sin[\theta]}{\sqrt{2}} & \frac{\sin[\theta]}{\sqrt{2}} \\ \frac{\sin[\theta]}{2} & \frac{-\cos[\theta] - e^{i\phi}}{2} & \frac{-\cos[\theta] + e^{i\phi}}{2} \\ -\frac{\sin[\theta]}{2} & \frac{-\cos[\theta] + e^{i\phi}}{2} & \frac{-\cos[\theta] - e^{i\phi}}{2} \end{bmatrix} \xrightarrow[\phi=0°]{\theta=90°} \begin{bmatrix} 0 & \frac{1}{\sqrt{2}} & -\frac{1}{\sqrt{2}} \\ \frac{1}{\sqrt{2}} & -\frac{1}{2} & \frac{1}{2} \\ -\frac{1}{\sqrt{2}} & \frac{1}{2} & -\frac{1}{2} \end{bmatrix}$$

A numerical simulation in HFSS is implemented and the device is shown in the Fig.4.

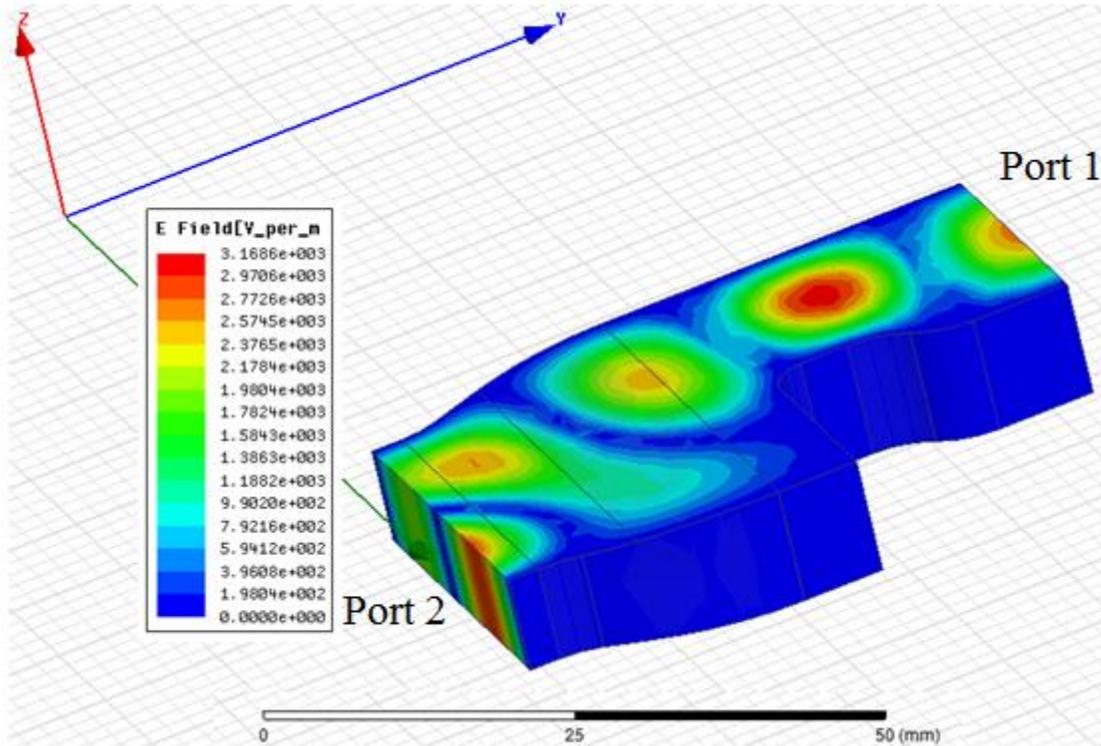

Fig.4. TE10 to TE10/TE20 converter and surface field pattern. The input power on WR 90 is 1 W.

The input port 1 supports the TE10 mode from RF source, and output E field pattern in port 2 is combination of TE10 and TE20 with 180 degree phase difference.  This converter starts with standard WR 90 rectangular waveguide for X band. First part of this converter is converting TE10 into TE10/TE20 over modes, while the second part is tapered the rectangular waveguide into the proper geometry to connect the input and output of rectangular waveguide in the center unit in the Fig. 3.  Double arc structure is utilized in the both transverse direction.  The scattering matrix of this three ports system is following in table 2.

Table.2: The scattering matrix of device in fig.4. Phase unit is degree..

| Freq | | S:1:1 | S:2:1 | S:2:2 |
|---|---|---|---|---|
| 11.424 (GHz) | 1:1 | (0.053789, 128) | (0.72844, -111) | (0.683, 17.6) |
| | 2:1 | (0.72844, -111) | (0.48473, 138) | (0.48417, 80.8) |
| | 2:2 | (0.683, 17.6) | (0.48417, 80.8) | (0.5469, 33.9) |

### 3.  Characteristic of the dual circular polarizer

After combining a center overmoded converter and two TE10 convertor at each end, the full geometry is shown in Fig 5.  Final scattering matrix is shown in table.2.

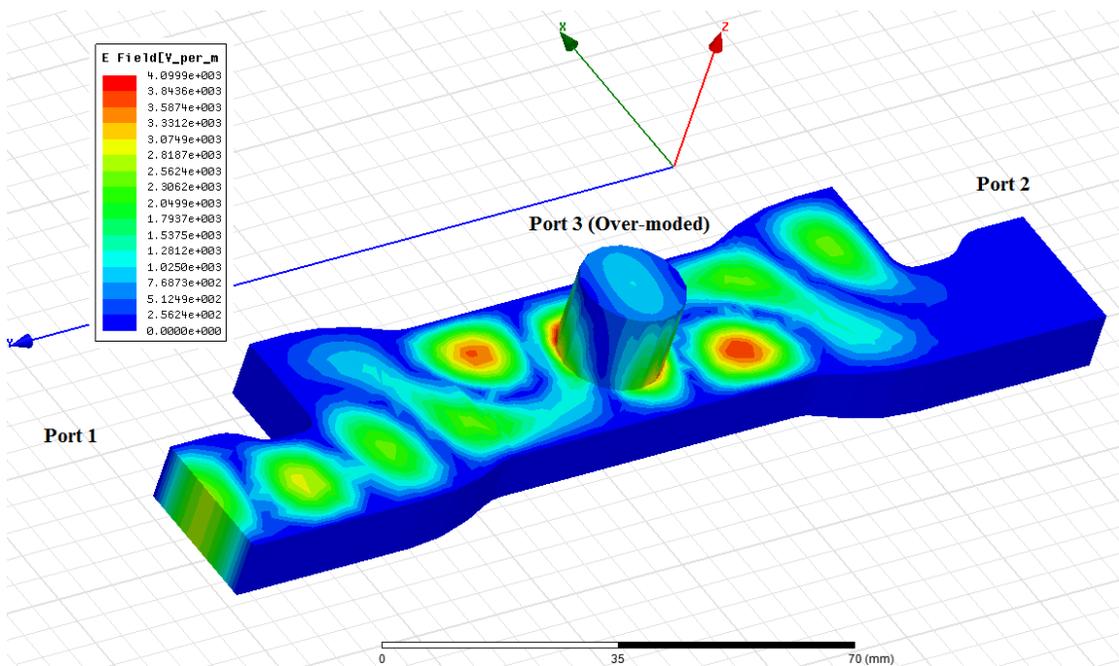

Fig.5. Final assembly of directional TE10 to dual polarizing circular converter. The input power on WR 90 is 1 W.

Table.2: The scattering matrix of device in fig.4. Units are decibel(dB) and degree.

| Freq | | S:1:1 | | S:1:2 | | S:2:1 | | S:2:2 | | S:3:1 | | S:3:2 | |
|---|---|---|---|---|---|---|---|---|---|---|---|---|---|
| 11.424 (GHz) | 1:1 | ( -17.5, | 120) | ( -29.9, | -33) | ( -23.4, | 105) | ( -50.5, | -143) | ( -3.19, | -92.7) | ( -3.03, | 109) |
| | 1:2 | ( -29.9, | -33) | ( -38.5, | -175) | ( -50.4, | 36.2) | ( -77.7, | 156) | ( -32.6, | -143) | ( -32.9, | 60.6) |
| | 2:1 | ( -23.4, | 105) | ( -50.4, | 36.2) | ( -17.2, | 119) | ( -29.9, | 147) | ( -3.19, | -92.6) | ( -3.03, | -70.7) |
| | 2:2 | ( -50.5, | -143) | ( -77.7, | 156) | ( -29.9, | 147) | ( -38.5, | -175) | ( -32.5, | 37.5) | ( -32.9, | 61.3) |
| | 3:1 | ( -3.19, | -92.7) | ( -32.6, | -143) | ( -3.19, | -92.6) | ( -32.5, | 37.5) | ( -13.9, | -120) | ( -51.3, | -79.9) |
| | 3:2 | ( -3.03, | 109) | ( -32.9, | 60.6) | ( -3.04, | -70.6) | ( -32.9, | 61.3) | ( -51.3, | -79.8) | ( -23, | -94) |

Considering the fact that there would be amplitude or phase errors in the circular output port, one ellipse converter is added to the output port of this converter to ensure two polarization TE11 modes have exact the same amplitude and 90 degree phase difference.

A section of ellipse waveguide has two different propagation constants at each polarization. For this reason, the two polarizations will generate a difference in phase delay which can compensate the phase difference from circular waveguide. Moreover, the match cone from circular to ellipse waveguides also has difference reflection coefficients, and the output can be matched equally even the input modes have different amplitude. The shorter cone will generate more higher order modes, thus, longer cone is preferred. This device has two exact two cones on the ends of the ellipse waveguide. There are three optimization parameters for a given set of amplitude and phase discrepancy and those three parameters are the lengths of cone and ellipse waveguide and the ellipse waveguide radius ratio. For our case, the design of ellipse converter is illustrated in Fig.6 and the scattering matrix is following in table 4.

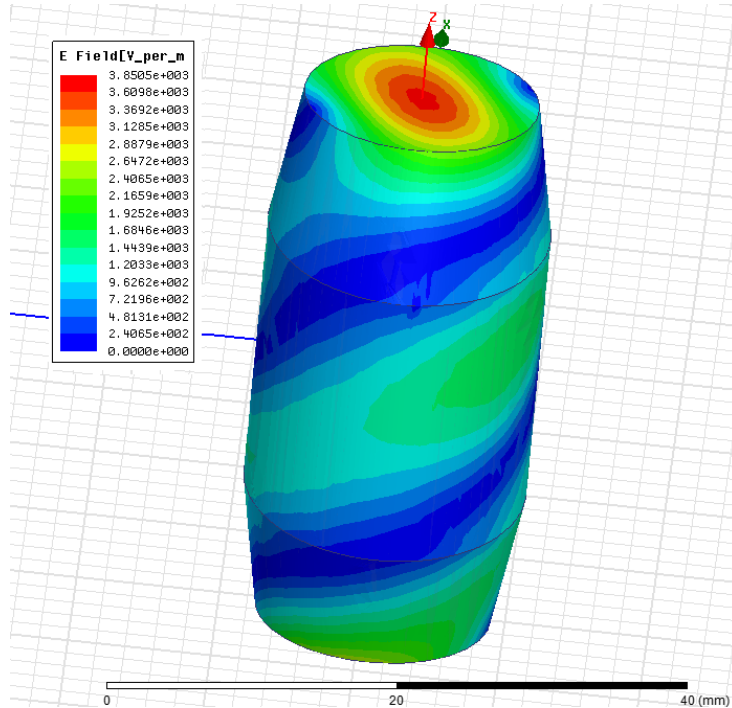

Fig.6. Ellipse converter and E field pattern on the surface. The power from input port (bottom) is 1 W for each polarization, and two polarizations are 90 degree apart.

Table.4: The scattering matrix of device in fig.6. Units are decibel(dB) and degree.

| Freq | | S:1:1 | | S:1:2 | | S:2:1 | | S:2:2 | |
|---|---|---|---|---|---|---|---|---|---|
| 11.424 (GHz) | 1:1 ( | -35.1, | -148) ( | -41.9, | 5.47) ( | -0.00163, | -11.3) ( | -59.7, | 94.5) |
| | 1:2 ( | -41.9, | 5.42) ( | -26.6, | -129) ( | -58.3, | 27.5) ( | -0.00982, | -58.8) |
| | 2:1 ( | -0.00163, | -11.3) ( | -56.6, | 15.7) ( | -35.1, | -54.3) ( | -41.9, | 104) |
| | 2:2 ( | -62.1, | 76.1) ( | -0.00983, | -58.8) ( | -41.9, | 104) ( | -26.6, | -168) |

If there is any amplitude or phase error due to the design or manufactory, this ellipse converter can be added to compensate either amplitude or phase. With the ellipse, the final design and E field on the surface are demonstrated in Fig.7.

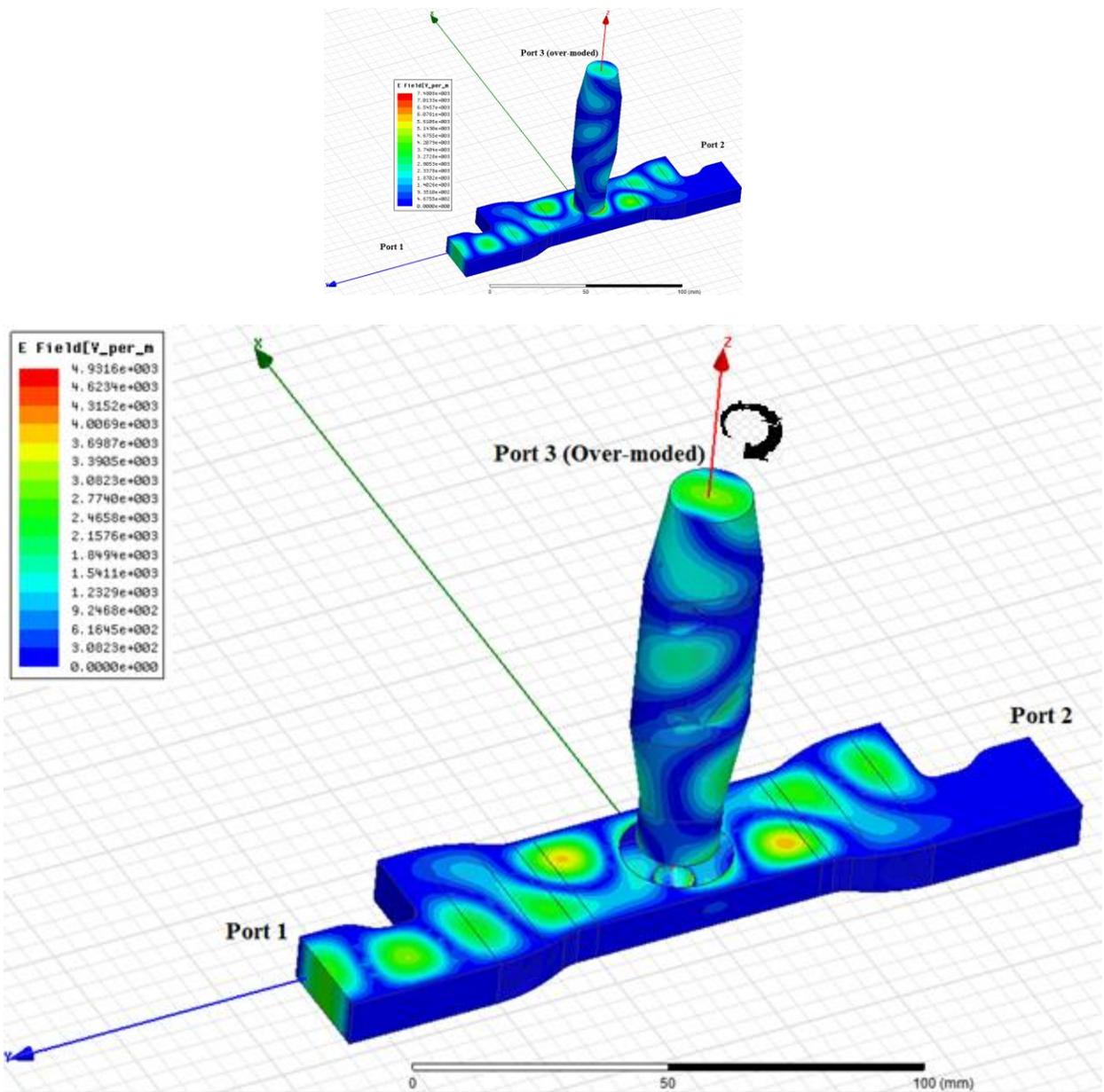

Fig.7. Final assembly of directional TE10 to dual polarizing circular converter with the ellipse converter. The input power on WR 90 is 1 W.

When the power input from one WR 90 rectangular waveguide is 1W, the max E field on the surface is around 7.4KV/m. If the input power is 50MW, the max E field on the surface is 52.3Mv/m. It is still less than the breakdown E field (~mV/m) for X band application.

Table.5: The scattering matrix of the final convert with ellipse corrector. Units are decibel(dB) and degree.

| Freq | S:1:1 | S:1:2 | S:2:1 | S:2:2 | S:3:1 | S:3:2 |
|------|-------|-------|-------|-------|-------|-------|
| 11.424 (GHz) 1:1 | ( -29.4, -11.2) | ( -29.4, -42.6) | ( -32, 13.6) | ( -54.4, -36.3) | ( -2.88, 14.7) | ( -3.16, 111) |
| 1:2 | ( -29.4, -42.6) | ( -39, -174) | ( -54.9, -32.6) | ( -80.5, -83) | ( -32.3, -33.6) | ( -33.1, 63.8) |
| 2:1 | ( -32, 13.6) | ( -54.9, -32.6) | ( -15.7, -37.3) | ( -28.6, -47.6) | ( -3.3, 20.9) | ( -2.96, -62.8) |
| 2:2 | ( -54.5, -36.4) | ( -80.5, -81.1) | ( -28.6, -47.6) | ( -38.9, -174) | ( -32.7, -27.5) | ( -32.9, -111) |
| 3:1 | ( -2.88, 14.7) | ( -32.3, -33.6) | ( -3.3, 20.9) | ( -32.7, -27.5) | ( -19.1, -118) | ( -23.2, -177) |
| 3:2 | ( -3.16, 111) | ( -33.1, 63.8) | ( -2.96, -62.8) | ( -32.9, -111) | ( -23.2, -177) | ( -21.5, 99.4) |

## 4. Another version of simplified polarizer.

A three ports converter in Fig. 7 has three physical ports, and there are two full standing wave pattern between two shorted planes. We can make one of the rectangular ports shorted. In the shorted structure, there are also two standing wave pattern. However, this design will only have two physical ports, and each physical port can support two modes. The two modes in the input port have two perfect matches with the modes in the circular output port respectively. Such a device would have a scattering matrix as following.

$$S = \begin{bmatrix} 0 & 0 & 1\times e^{j\theta_1} & 0 \\ 0 & 0 & 0 & 1\times e^{j\theta_2} \\ 1\times e^{j\theta_1} & 0 & 0 & 0 \\ 0 & 1\times e^{j\theta_2} & 0 & 0 \end{bmatrix}$$

Where $\theta_1$ and $\theta_2$ are the phase delay for TE10 and TE20 modes in this structure.

This geometry can be also understood as a rectangular waveguide with one short end, and on the center top of the plane, there is a circular coupler to extract two polarization of TE11 mode. TE10 and TE20 are both trapped as standing wave and form TE010 and TE020 modes. The maximums H fields have the same magnitude and occur at the same location where the circular waveguide are. Our design goal is achieving maximum extraction and the extracted two polarizations have 90 degree phase delay and the same magnitude. To realize that, TE10 and TE20 mode is matched individually. The distance from the center of circular waveguide to the short end for each mode may not be the same length. One fin groove is added to match both TE10 and TE20 modes. For TE10 mode, the plane where the fin acts like short end to TE10, because TE10 is evanescent mode inside of the fin end. However TE20 can prorogate with little reflection in the fin area. Length of the fin is related to the difference of the TE10/TE20 propagation constant. The design geometer of such rectangular to circular waveguide is given in Fig. 8.

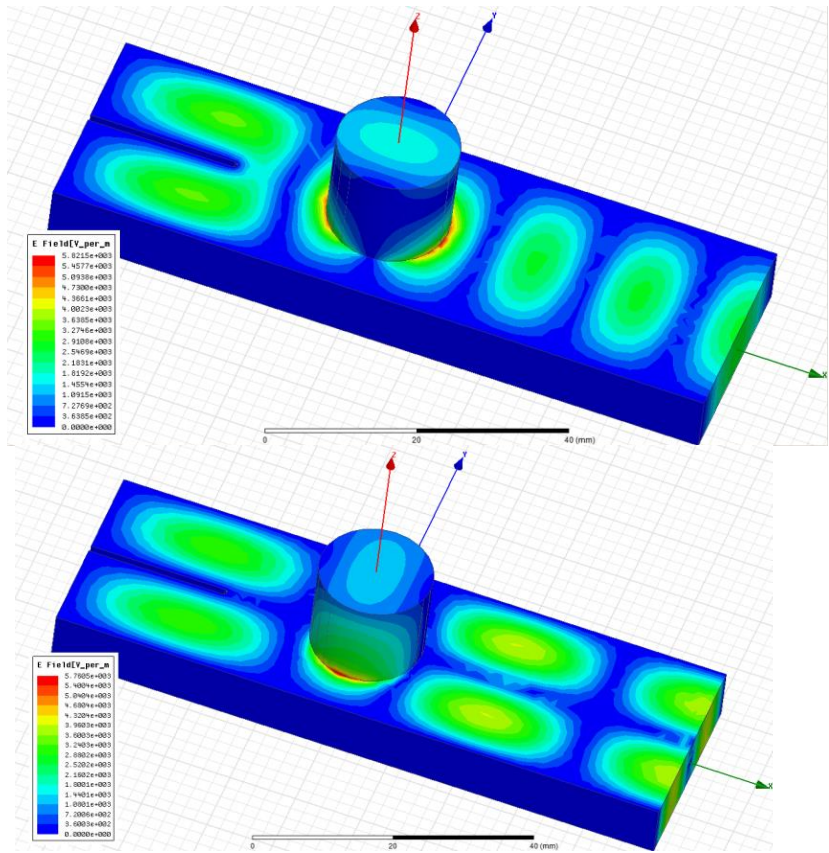

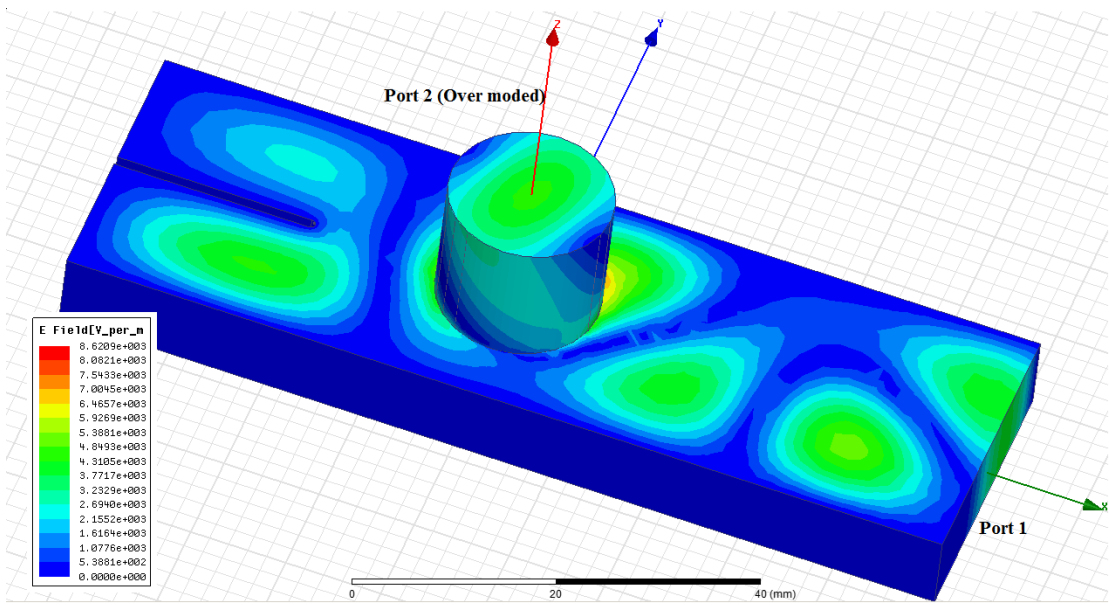

Fig.8: This is the E field pattern on the mode converter. a: TE10 converter; b: TE10 converter; c: over-moded converter with arbitrary phase delay (30° in the figure). The input power on port1 is 1 W for TE10/TE20 modes individually.

Similarly to Fig. 3, the output combination of circular waveguide is one TE11 mode rotating around the center axis. The 90 degree phase difference is adjustable by the input TE10 and TE20 phase difference. Each individual mode phase delay $\theta_1$ and $\theta_2$ in this converter add the input phase difference between TE10 and TE20 shall give a total phase difference, and that number can be adjusted to 90 degree for two polarizations. The input TE10 and TE20 phase difference can be achieved by an additional length of overmoded rectangular waveguide in which TE10 and TE20 modes have difference propagation constants. The scattering matrix parameter is given in table 6.

Table.6: The scattering matrix of device in fig.8. Units are decibel(dB) and degree.

| Freq | | S:1:1 | S:1:2 | S:2:1 | S:2:2 |
|---|---|---|---|---|---|
| 11.424 (GHz) | 1:1 | ( -21, 70.3) | ( -66.6, -77.9) | ( -0.0344, 65.5) | ( -69.9, -32.8) |
| | 1:2 | ( -66.6, -77.9) | ( -21.1, 21.9) | ( -68.8, 28.8) | ( -0.0337, 110) |
| | 2:1 | ( -0.0344, 65.5) | ( -68.8, 29.7) | ( -21, -119) | ( -66.8, 74.1) |
| | 2:2 | ( -70, -31.8) | ( -0.0337, 110) | ( -66.8, 74.1) | ( -21.1, 17.6) |

The converter to convert TE10 to TE10/TE20 modes is firstly designed by Chris Nantista. However, we optimize it to address our application. A device in Fig. 7 can convert TE10 from one arm waveguide into TE10/TE20 combination modes in the center body without leakage to the other arm. The power from WR 90 input is equally splitted into TE10/TE20 modes in the over-moded waveguide in the center. If circular waveguide is shorted, the power will output to the second arm without leakage to the first arm. This characteristic can be applied to an X band directional waveguide converter. The waveguide geometry and E field pattern on the surface are shown in Fig. 9, and followed by its scattering matrix in table 7.

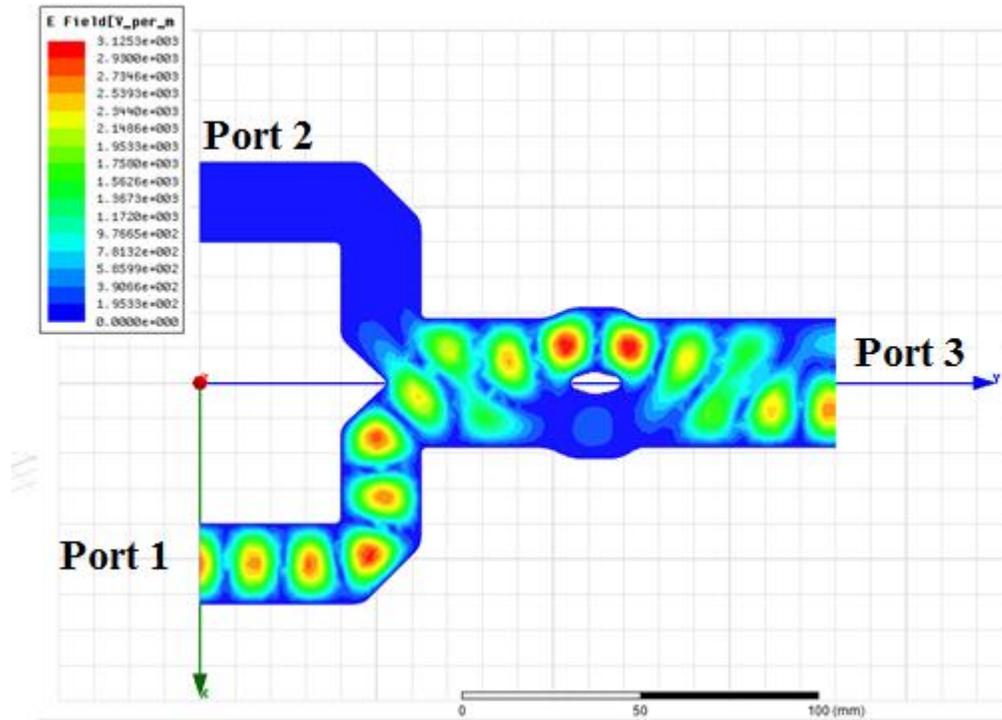

Fig.9: This is the E field pattern on the mode converter. The input power on port1 is 1 W.

Table.7: The scattering matrix of converter in fig.9. Units are decibel (dB) and degree.

| Freq | | S:1:1 | | S:2:1 | | S:3:1 | | S:3:2 | |
|---|---|---|---|---|---|---|---|---|---|
| 11.424 (GHz) | 1:1 ( | -19.4, | 106) ( | -27.4, | 165) ( | -3.13, | 160) ( | -3.01, | 175) |
| | 2:1 ( | -27.4, | 165) ( | -17.6, | 101) ( | -3.08, | 160) ( | -3.11, | -5.84) |
| | 3:1 ( | -3.13, | 160) ( | -3.08, | 160) ( | -16.8, | 22.2) ( | -37.6, | -108) |
| | 3:2 ( | -3.01, | 175) ( | -3.11, | -5.84) ( | -37.6, | -108) ( | -19.5, | 86.4) |

By cascading these two structures, the second version of X-band dual polarization converter assembly is shown in the Fig. 10. Similarly, when the power input from one WR 90 rectangular waveguide is 1W, the max E field on the surface is around 5000V/m. If the input power is 50MW, the max E field on the surface is 35MV/m. It is still less than the typical breakdown E field (200mV/m) for X band application. The geometry is shown in Fig. 10 and followed by a 4×4 scattering matrix in table 8.

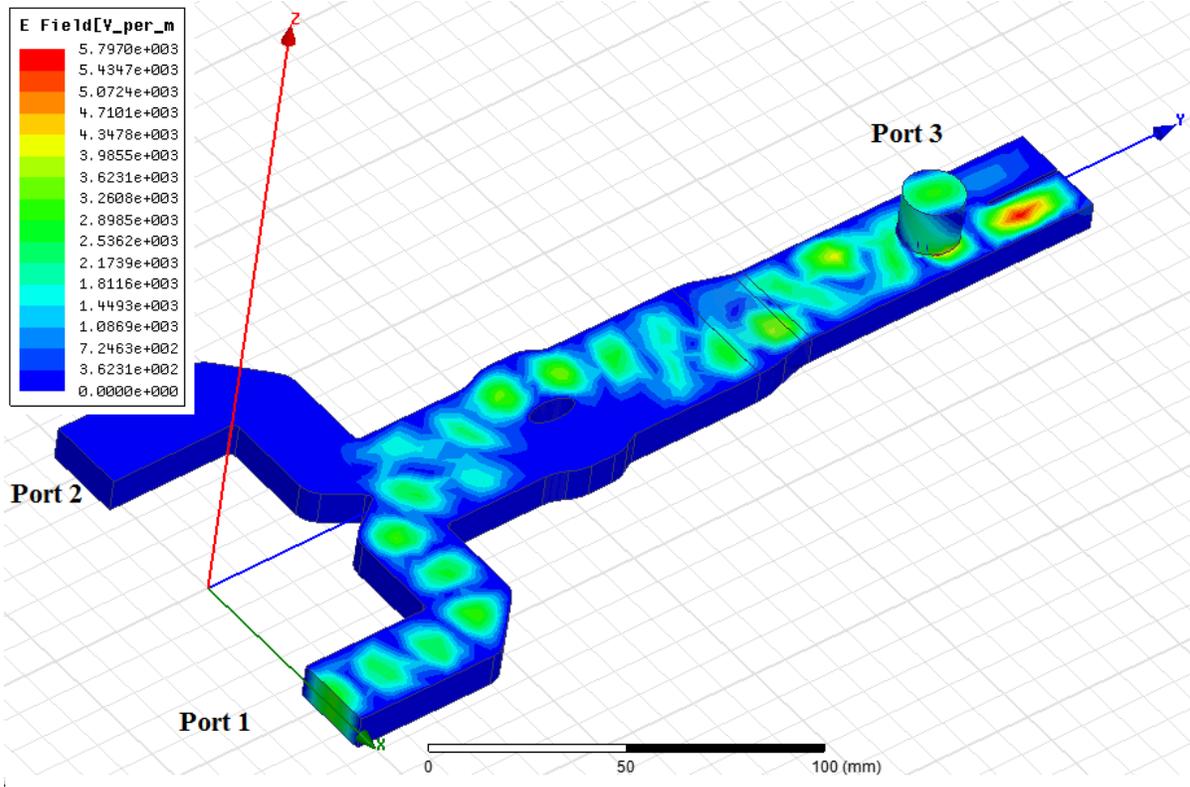

Fig.10. Final assembly of directional TE10 to dual polarizing circular converter and the complex E field magnitude is plotted. The input power on WR 90 is 1 W and max E field is 7.4 KV/m.

Table.8: The scattering matrix of second version in fig.10. Units are decibel (dB) and degree.

| Freq | | S:1:1 | | S:2:1 | | S:3:1 | | S:3:2 | |
|---|---|---|---|---|---|---|---|---|---|
| 11.424 (GHz) | 1:1 ( | -37.9, | -57.3) ( | -45.6, | 89) ( | -3, | -21.4) ( | -3.02, | 69.4) |
| | 2:1 ( | -45.6, | 89) ( | -35, | -30.4) ( | -3.02, | -21.6) ( | -3, | -111) |
| | 3:1 ( | -3, | -21.4) ( | -3.02, | -21.6) ( | -38.3, | 159) ( | -47.1, | 36.3) |
| | 3:2 ( | -3.02, | 69.4) ( | -3, | -111) ( | -47.1, | 36.3) ( | -34.6, | 12.2) |

## 5. Frequency response, Transient response and Thermal concern.

### 5.1. Frequency response:

Both versions of dual mode polarizers have broad operational frequency ranges. The scattering parameters as a function of frequency are scanned and illustrated in fig 11.

From fig.12, even the amplitude of scattering parameters are acceptable within certain bandwidth, however the phase difference of two TE11 modes has broad bandwidth around 50MHz. This shall be overcome by the ellipse phase shifter in Fig 6.

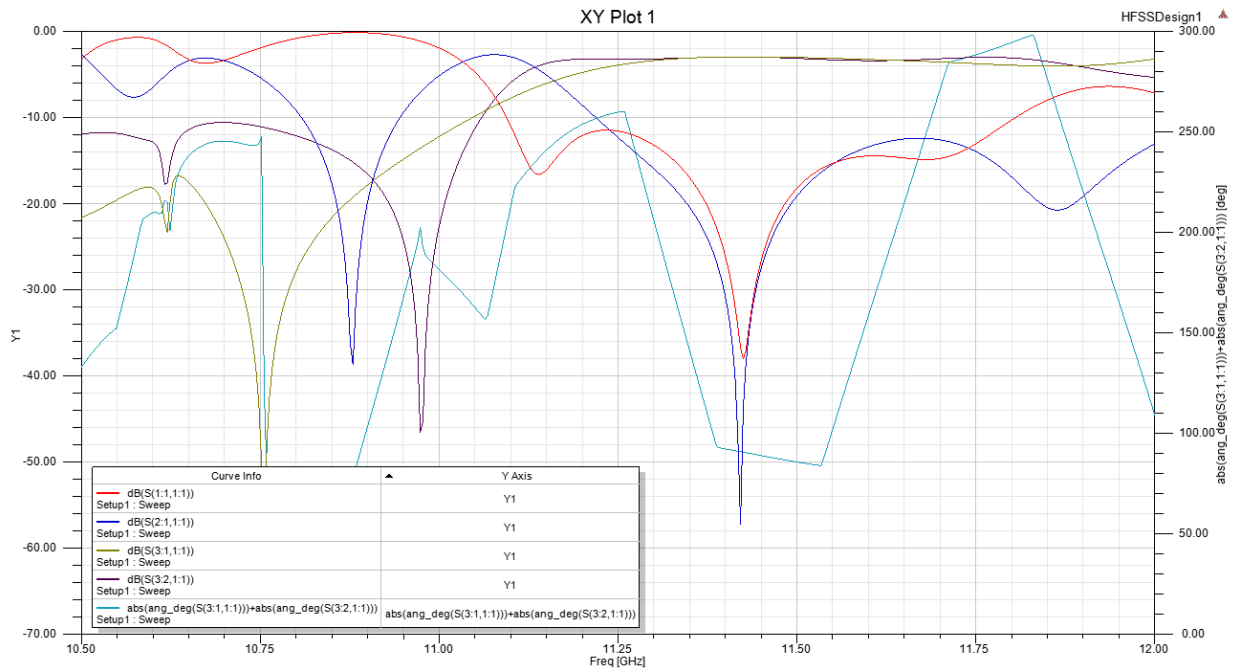

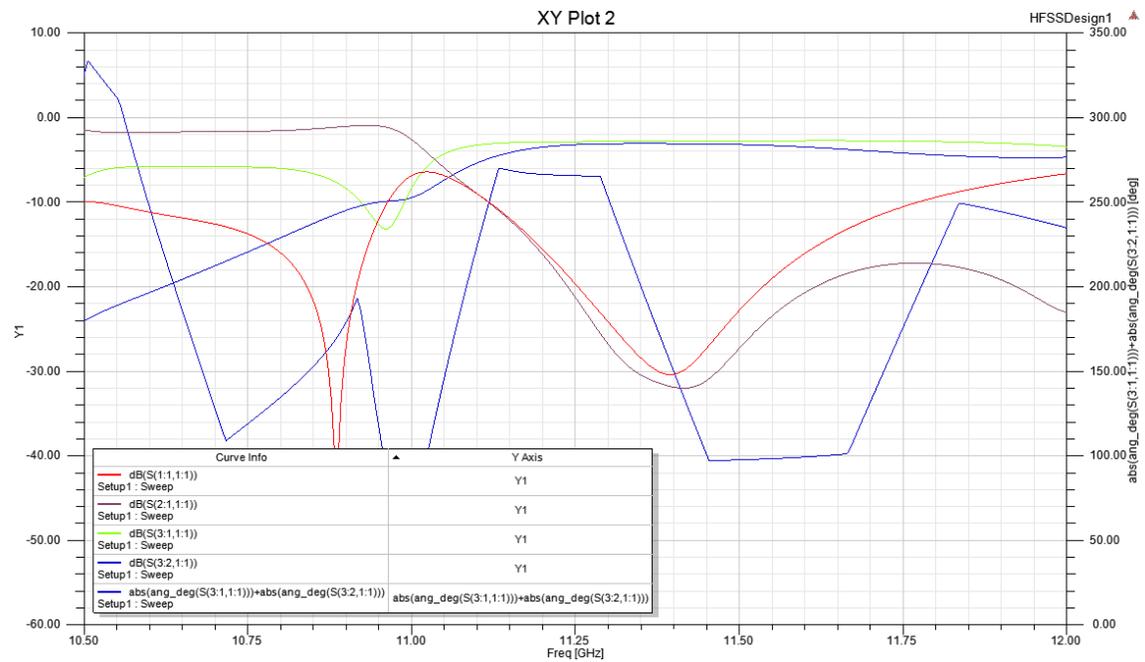

Fig. 12. Frequency spectrum of the scattering parameter of the dual polarizers.

The complex magnitude of E field over a RF cycle on the surface of two different versions polarizer with different frequencies are shown in fig. 13.

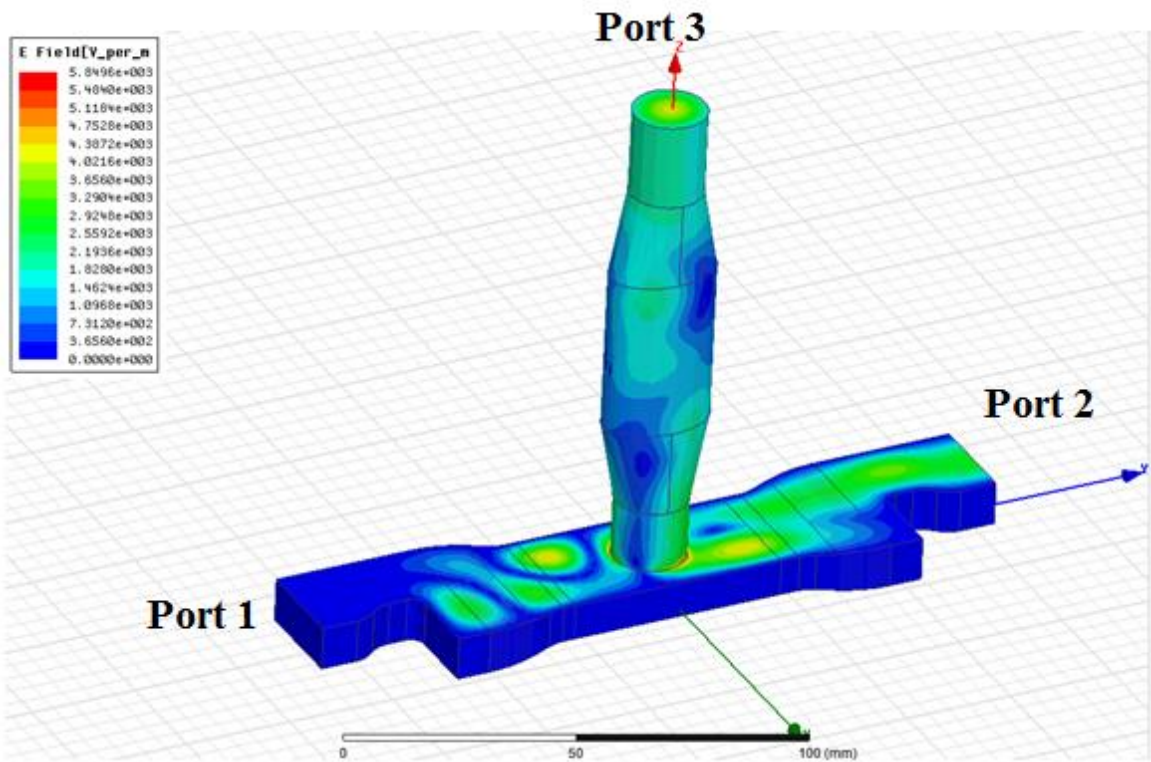

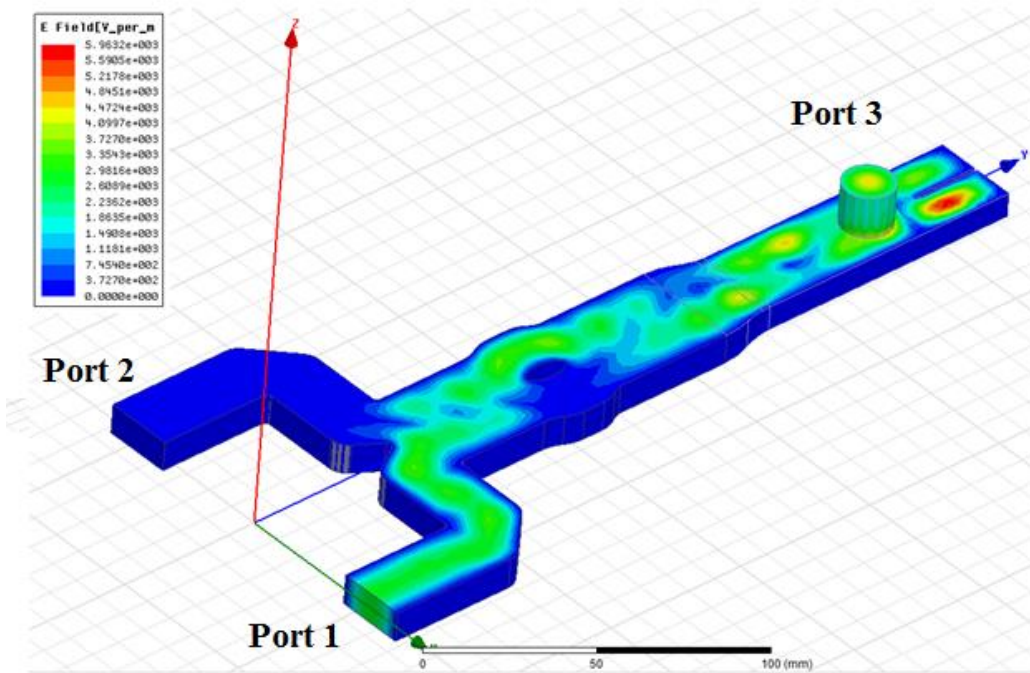

Fig 13. The magnitude of surface E field over a RF cycle is plotted on two versions of X band polarizers.

### 5.2. Transient response: (don't want to include, ask for opinions)

An x band klystron usually gives a pulse with duration around 1.5 µs and a rising time of 15 ns. The main lobe of spectrum has a bandwidth of 1.3MHz, while the side lubes are spacing 0.7MHz apart. The spectrum is shown fig 13.

Fig.13. Time domain and frequency spectrum from a square pulse from a klystron .

Convoluted with the S3:1,1:1 and S3:2,1:1 in the frequency domain, a transient pulse responses of two different modes are obtained from this polarizer. The Responses to the pulse for each polarizer are given in fig. 14.

Fig.14. Final assembly of directional TE10 to dual polarizing circular converter. The input power on WR 90 is 1 W and max E field is 7.4 KV/m.

Note, if polarizers received a dual circular polarized pulse from the circular waveguide, the output on port 3 should be the same as shown in fig. 14.

### 5.3. Thermal concern:

These two converters are designed to deliver multi-megawatts RF power into dual polarization TE11 modes. Thus, the power loss on the surface can be non-trivial and required consideration. Water cooling pipes are considered to keep the whole system exterior around room temperature. Fig 15 below shows the power loss maps of each system under the condition that input power is 1 W.

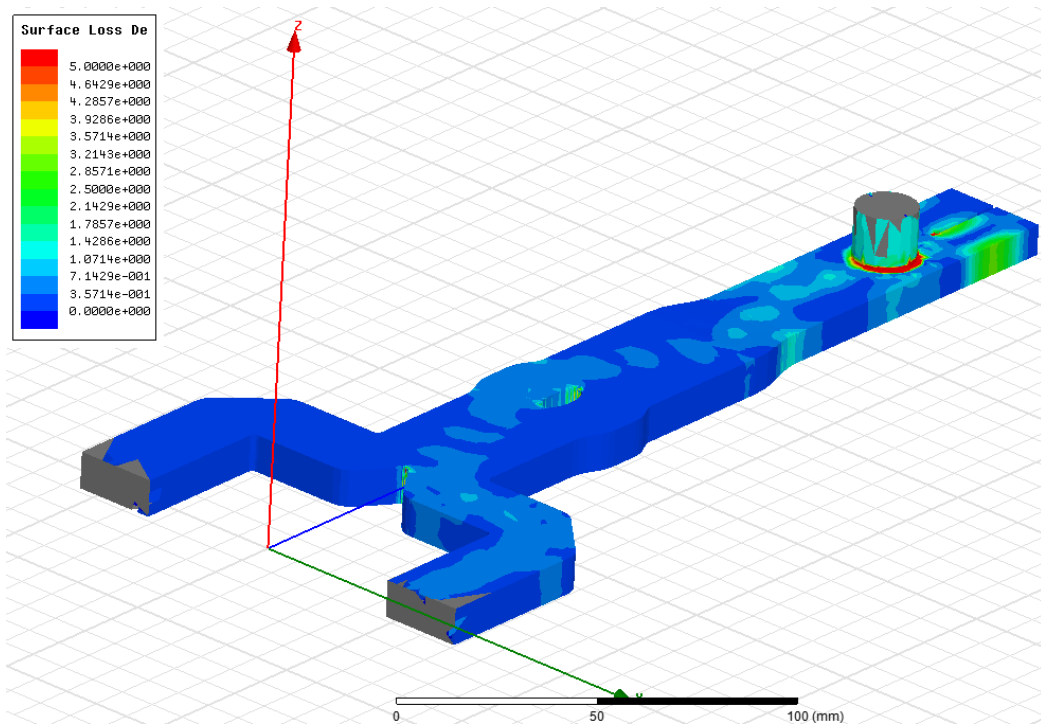

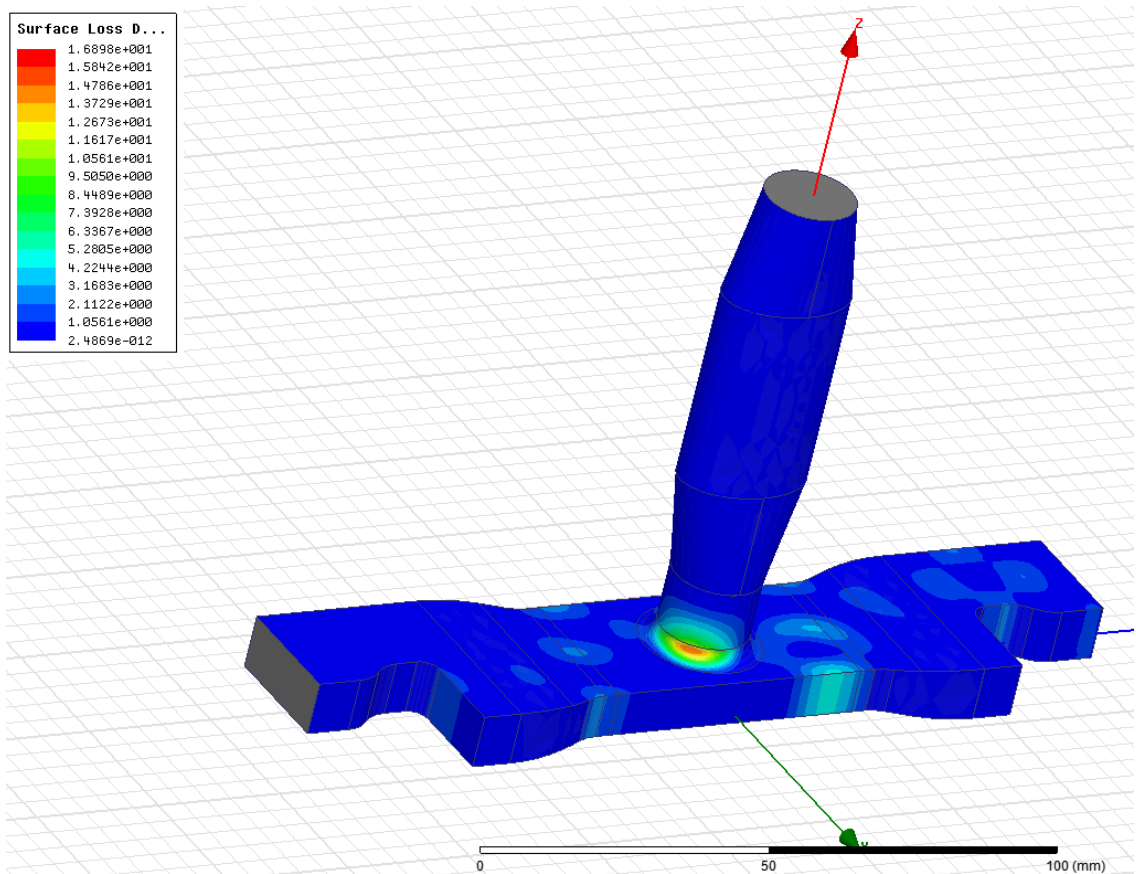

Fig. 15. Snapshot on surface RF loss distribution in a RF cycle for each version of the dual polarizers.

The main RF power loss concentrates on bottom of the circular waveguide, even though there is ring chamfer in order to reduce the local H field. The integration RF losses over a RF cycle for each system from fig 15 are 0.0117W and 0.0108W respectively when the input power is 1W. The total loss of these two device is around 1% input RF power. Temperature dependency of copper electric conductivity will be including in future study.

## 6. Conclusion

We present two X band novel dual circular polarization over-moded converter. The design used all simple shapes and it will be easy to manufacture. This polarizer can be utilized for several of motivations, including detector of CMB, high power RF load and RF power compressor.

## Acknowledgement

We would like to thank Dr Chris Nantista from SLAC national accelerating laboratory for some useful discussions. This work was supported by Department of Energy Contract No. DE-AC02-76SF00515.

## Reference

1. H. Padamsee, J. Knobloch, and T. Hays, RF Superconductivity for Accelerators. 2nd Edition (Wiley and Sons, New York,NY, 2008).
2. W.-D. Moeller for the TESLA Collaboration, "High Power Coupler For The TESLA Test Facility", Proceedings of the 9th Workshop on the RF Superconductivity, 1999, Santa Fe, V.2, pp.577-581.
3. J. R. Delayen, L.R. Doolittle, T. Hiatt, J. Hogan, J. Mammosser. "An R.F. Input Coupler System For The CEBAF Energy Upgrade Cryomodule." Proceedings of the 1999 Particle Accelerator Conference, New York, 1999. pp1462-1464.
4. S. Belomestnykh, et al., "High Average Power Fundamental Input Couplers for the Cornell University ERL: Requirements, Design Challenges and First Ideas," Cornell LEPP Report ERL 02-8 (September 9, 2002).
5. RF windows
6. E. Snitzer. "Cylindrical dielectric waveguide modes" Journal of the Optical Society of America, Vol. 51, Issue 5, pp. 491-498 (1961)
7. Zaki, K.A. Atia, A.E. "Modes in Dielectric-Loaded Waveguides and Resonators." Microwave Theory and Techniques, IEEE Transactions on V.31 , I.12 ,1982. pp1039 – 1045.
8. Clarricoats, P.J.B ."Properties of dielectric-rod junctions in circular waveguide", Electrical Engineers, Proceedings of the Institution of (Volume:111 , Issue: 1 ). 1964. Pp.43 – 50
9. Clarricoats, P.J.B. ; Taylor, B.C. "Evanescent and propagating modes of dielectric-loaded circular waveguide". Electrical Engineers, Proceedings of the Institution of  (Volume:111 ,  Issue: 12 ) 1964 pp.1951-1956.
10. Walter M. Elsasser. "Attenuation in a Dielectric Circular Rod". J. Appl. Phys. 20, 1193 (1949)

11. Rothwell, E.J. and Frasch, L.L. "Propagation characteristics of dielectric-rod-loaded waveguides".Microwave Theory and Techniques, IEEE Transactions on  (Volume:36 ,  Issue: 3 ) 1988. pp 594 – 600.
12. Optics book
13. S.F.Mahmoud. "Electromagnetics waveguides theory and applications". The Institution of Engineering and Technology (December 1991)
14. A.A. Mostafaa, C.M. Krowneb, K.A. Zakic & S. Tantawid  "Hybrid-Mode Fields in Isotropic and Anisotropic Planar Microstrip Structures." Journal of Electromagnetic Waves and Applications Volume 5, Issue 6, 1991

**Appendix:**

$$S = \begin{bmatrix} \dfrac{-\cos[\theta_{01}] - e^{i\phi_{01}}}{2} & 0 & \dfrac{-\cos[\theta_{01}] + e^{i\phi_{01}}}{2} & 0 & \dfrac{\sin[\theta_{01}]}{\sqrt{2}} & 0 \\ 0 & \dfrac{-\cos[\theta_{02}] - e^{i\phi_{02}}}{2} & 0 & \dfrac{-\cos[\theta_{02}] + e^{i\phi_{02}}}{2} & 0 & \dfrac{\sin[\theta_{02}]}{\sqrt{2}} \\ \dfrac{-\cos[\theta_{01}] + e^{i\phi_{01}}}{2} & 0 & \dfrac{-\cos[\theta_{01}] - e^{i\phi_{01}}}{2} & 0 & \dfrac{\sin[\theta_{01}]}{\sqrt{2}} & 0 \\ 0 & \dfrac{-\cos[\theta_{02}] + e^{i\phi_{02}}}{2} & 0 & \dfrac{-\cos[\theta_{02}] - e^{i\phi_{02}}}{2} & 0 & \dfrac{\sin[\theta_{02}]}{\sqrt{2}} \\ \dfrac{\sin[\theta_{01}]}{\sqrt{2}} & 0 & \dfrac{\sin[\theta_{01}]}{\sqrt{2}} & 0 & \cos[\theta_{01}] & 0 \\ 0 & \dfrac{\sin[\theta_{02}]}{\sqrt{2}} & 0 & \dfrac{\sin[\theta_{02}]}{\sqrt{2}} & 0 & \cos[\theta_{02}] \end{bmatrix}$$

Table 5

| Freq | | S:1:1 | | S:2:1 | | S:3:1 | | S:3:2 | |
|---|---|---|---|---|---|---|---|---|---|
| 11.424 (GHz) | 1:1 ( | -24.8, | -155) ( | -24.9, | 171) ( | -3.17, | 136) ( | -2.91, | -142) |
| | 2:1 ( | -24.9, | 171) ( | -14.6, | -102) ( | -3.02, | -35.2) ( | -3.35, | -134) |
| | 3:1 ( | -3.17, | 136) ( | -3.02, | -35.2) ( | -18.8, | -168) ( | -22.8, | 101) |
| | 3:2 ( | -2.91, | -142) ( | -3.35, | -134) ( | -22.8, | 101) ( | -16.7, | 44.4) |